\documentclass[aps,prl,twocolumn,groupedaddress,bibnotes,citeautoscript]{revtex4-1}

\usepackage{graphicx}
\usepackage{dcolumn}
\usepackage{bm}
\usepackage{color}

\begin{document}


\title{Interband and polaronic excitations in YTiO$_3$ from first principles}

\author{Burak Himmetoglu, Anderson Janotti, Lars Bjaalie, and Chris G. Van de Walle}
\affiliation{Materials Department, University of California, Santa Barbara, CA 93106-5050}

\date{\today}

\begin{abstract}
YTiO$_3$, as a prototypical Mott insulator, has been the subject of numerous experimental investigations of its electronic structure.
The onset of absorption in optical conductivity measurements has generally been interpreted to be due to interband transitions at the
fundamental gap.  Here we re-examine the electronic structure of YTiO$_3$ using
density functional theory with either a Hubbard correction (DFT+$U$) or a hybrid functional.
Interband transitions turn out to be much higher in energy than the observed onset of
optical absorption. However, in case of $p$-type doping, holes tend to become self-trapped
in the form of small polarons, localized on individual Ti sites.
Exciting electrons from the occupied lower Hubbard band to the small-polaron state then leads to broad infrared
absorption, consistent with the onset in the experimental optical conductivity spectra.
\end{abstract}


\maketitle

Rare-earth titanates (RTiO$_3$) are typical Mott insulators,
characterized by strong electronic correlations that
dominate their electronic structure~\cite{Mott,Imada-rev}.
The insulating gap is formed between the lower and upper Hubbard bands
(LHB and UHB) which are mainly derived from Ti $d$ orbitals.
Electronic correlations favor integer occupations of Ti-$d$-derived
bands, leading to magnetic ordering at low temperatures.
The details of the magnetic ordering depend critically on
structural distortions
~\cite{kanamori-sx,goodenough-rev},
which results in a remarkable interplay between electronic, magnetic, and structural degrees of freedom.

A pronounced interest in these materials has emerged
in the area of complex oxide interfaces~\cite{hwang-emergent},
with the realization of a two-dimensional electron gas (2DEG)
in SrTiO$_3$/LaAlO$_3$ heterostructures~\cite{ohtomo-2deg}.
Record-high 2DEG densities of the order 3$\times$10$^{14}$ cm$^{-2}$
have recently been achieved at the SrTiO$_3$/GdTiO$_3$
interface~\cite{moetakef-2deg}. While the first
interface is formed between two band insulators, the
latter involves a Mott insulator, which
calls for a deeper understanding of electronic
structure of Mott insulating rare-earth titanates in general.

\begin{figure}[!ht]
\includegraphics[width=0.40\textwidth]{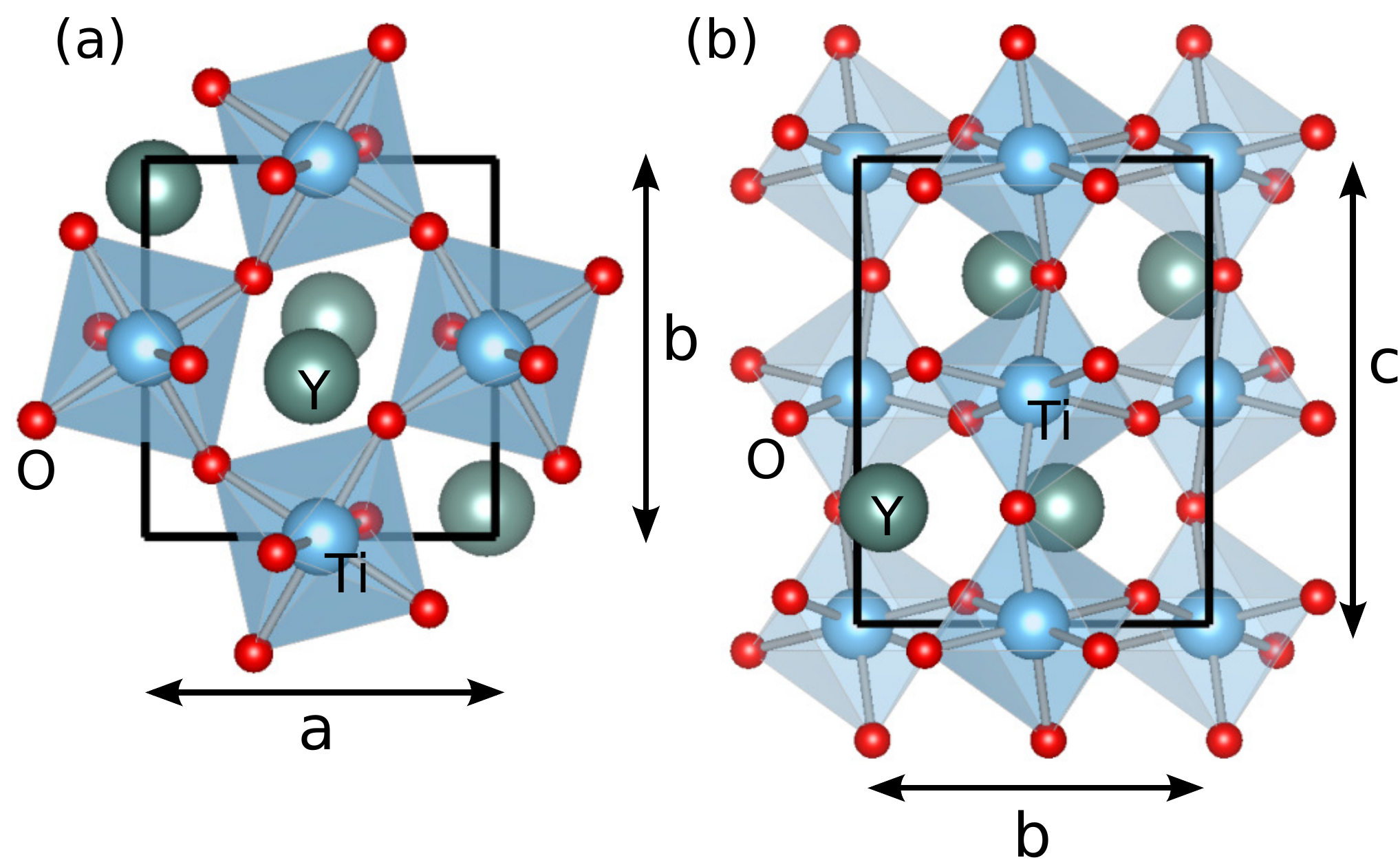}
\caption{\label{fig:structure}(color online) Primitive cell of YTiO$_3$.
(a) Top view showing the $a$ and $b$ lattice parameters. (b) Side
view showing the $b$ and $c$ lattice parameters.}
\end{figure}

Here we investigate the electronic and structural properties of the rare-earth titanate
YTiO$_3$, a representative Mott-insulator complex oxide.
It assumes an orthorhombic perovskite-type crystal structure (space group $Pbnm$),
with 20 atoms in the primitive cell, as represented in Fig.~\ref{fig:structure}.
Although this compound has been the subject of numerous experimental~\cite{kovaleva-opt,gossling-opt,okimoto-opt,arita-ipes,morikawa-pes}
and theoretical~\cite{pavarini-njp,pavarini-dmft,craco-dmft} studies,
its electronic properties are far from fully understood.
The band gap is usually deduced from optical conductivity spectra,
which have a tail in the infrared region ~\cite{kovaleva-opt,gossling-opt,okimoto-opt}.
The onset around 0.6 eV has been attributed to interband
transitions associated with a small band gap of 0.6-1.0 eV.
In order to elucidate these experimental findings, we have investigated the
electronic structure of YTiO$_3$ using first-principles calculations.
We employ density functional theory (DFT) with a Hubbard-corrected exchange-correlation
functional (DFT+$U$), and also with a hybrid functional. Both approaches consistently yield a
band gap significantly larger than the measured onset of optical absorption, casting severe doubt on
the attribution to transitions across the fundamental gap.

Instead we propose here that the onset of the conductivity spectra is due to excitations related to small polarons.
The presence of hole polarons is consistent with the observation of unintentional $p$-type doping and the
temperature dependence of resistivity in rare-earth titanates~\cite{goodenough-p}.
Our first-principles studies indeed show that self-trapped holes are stable in the form of small polarons,
and that optical absorption associated with these
polarons can explain the low-energy absorption onset.
We note that studying YTiO$_3$ also sheds light on the electronic
properties of GdTiO$_3$ \cite{moetakef-2deg},
which has structural and electronic properties very similar to YTiO$_3$, but for which scant
experimental information is available.

Standard exchange-correlation functionals generally yield
a poor description of the electronic structure of strongly
correlated systems. The failure is particularly pronounced in the case of
partially filled bands derived from localized transition-metal $d$ orbitals, where
it leads to significant errors in ground-state properties, magnetic order, and band gaps.
These shortcomings can be overcome by the use of
exchange-correlation functionals that incorporate a corrective term
inspired by the Hubbard Model (DFT+$U$), or with hybrid functionals that
include a fraction of exact exchange.
We use both approaches in our study.

The DFT+$U$ calculations are performed with the plane-wave self-consistent field
(PWSCF) code of the {\it Quantum Espresso}
package~\cite{QE}, using ultra-soft pseudopotentials~\cite{uspp}.
We use the generalized gradient approximation (GGA) of
Perdew, Burke, and Ernzerhof (PBE) ~\cite{pbe}.
The electronic wavefunctions and charge density are expanded up to kinetic energy cutoffs of
$50\, {\rm Ry}$ and $600\, {\rm Ry}$, respectively.
The parameter $U$ acting on Ti $d$ orbitals is calculated self-consistently
using linear response~\cite{ucalc-1,ucalc-2}, resulting in $U$=3.70 eV.
We have also considered the possibility of including Hund's coupling $J$
in the DFT+$U$ corrective functional, and found that it
does not lead to considerable differences (see Sec.\ I of Supplemental Material \cite{sup} for details).

For the hybrid functional calculations, we adopt the Heyd-Scuseria-Ernzherof (HSE) functional~\cite{hse-1,hse-2},
with a mixing parameter of $25\%$ and screening parameter of $0.2$ \AA.
The HSE calculations are performed using the projected-augmented wave (PAW) method as implemented
in the VASP code~\cite{vasp-1,vasp-2}. An energy cut-off of 400 eV for
the expansion of plane waves is used. For Brillouin-zone integrations,
a 4$\times$4$\times$4 special-point grid is used for both DFT+$U$ and HSE calculations.
Full structural relaxations are always performed.

The calculated orthorhombic structure for bulk YTiO$_3$, displaying GdFeO$_3$-type distortions,
is in agreement with experiment~\cite{yto-structure}; see
Table~\ref{tab:structure}. DFT+$U$ overestimates
the $a$ and $b$ lattice parameters by 1.5\%, and $c$ by 3\%, in line with the expected accuracy of that methodology.
HSE yields a higher accuracy, with an underestimation of $a$ and $b$ by only 0.5\%, and $c$ by only 1\%.

\begin{table}[!ht]
\caption{\label{tab:structure}  Lattice parameters $a$, $b$, and $c$ (\AA)
for YTiO$_3$ in the ground-state orthorhombic crystal structure calculated with DFT+$U$ and with the HSE hybrid functional.
Experimental values from Ref.~\onlinecite{yto-structure} are listed for comparison.}
\begin{ruledtabular}
\begin{tabular}{cccc}
 lattice parameter & DFT+$U$ & HSE & Exp. \\
\hline
$a$  & 5.40 & 5.29 & 5.33 \\
$b$  & 5.76 & 5.70 & 5.68 \\
$c$  & 7.85 & 7.53 & 7.61
\end{tabular}
\end{ruledtabular}
\end{table}

The electronic band structure and atom-projected densities of states (DOS)
resulting from the HSE calculations are shown in Fig.~\ref{fig:hse-bands}.
The band structure resulting from the DFT+$U$ calculation
(see Sec.\ I of Supplemental Material \cite{sup} ) is in overall
agreement with HSE.
%
HSE results in a band gap of $E_g$=2.07 eV, while
DFT+$U$ results in $E_g$=2.20 eV (error bars are discussed in Sec. I of Ref.~\onlinecite{sup}).
The LHB and UHB are rather flat due to the localized nature of the Ti $d$ orbitals.
A previous study using a modified HSE functional with a correlation potential based on the
local density approximation (LDA), instead of PBE, reported a band gap of 1.41 eV for YTiO$_3$~\cite{rubio}.
While this gap is somewhat smaller than ours, it is still considerably larger than the onset of optical conductivity.
The valence-band maximum is between $Z$ and $T$ points
and the conduction-band minimum is at $\Gamma$.
As can be seen in Fig.~\ref{fig:hse-bands}(a),
the LHB consists entirely of spin-up states while
spin-down bands are higher in energy, thus
YTiO$_3$ displays a ferromagnetic ordering at low temperature,
in agreement with experiments that reported a Curie temperature of $T_{\rm C}$=29 K~\cite{Goral}.
The magnetic moments originate from the Ti atoms in a +3 oxidation state.


\begin{figure}[!ht]
\includegraphics[width=0.40\textwidth]{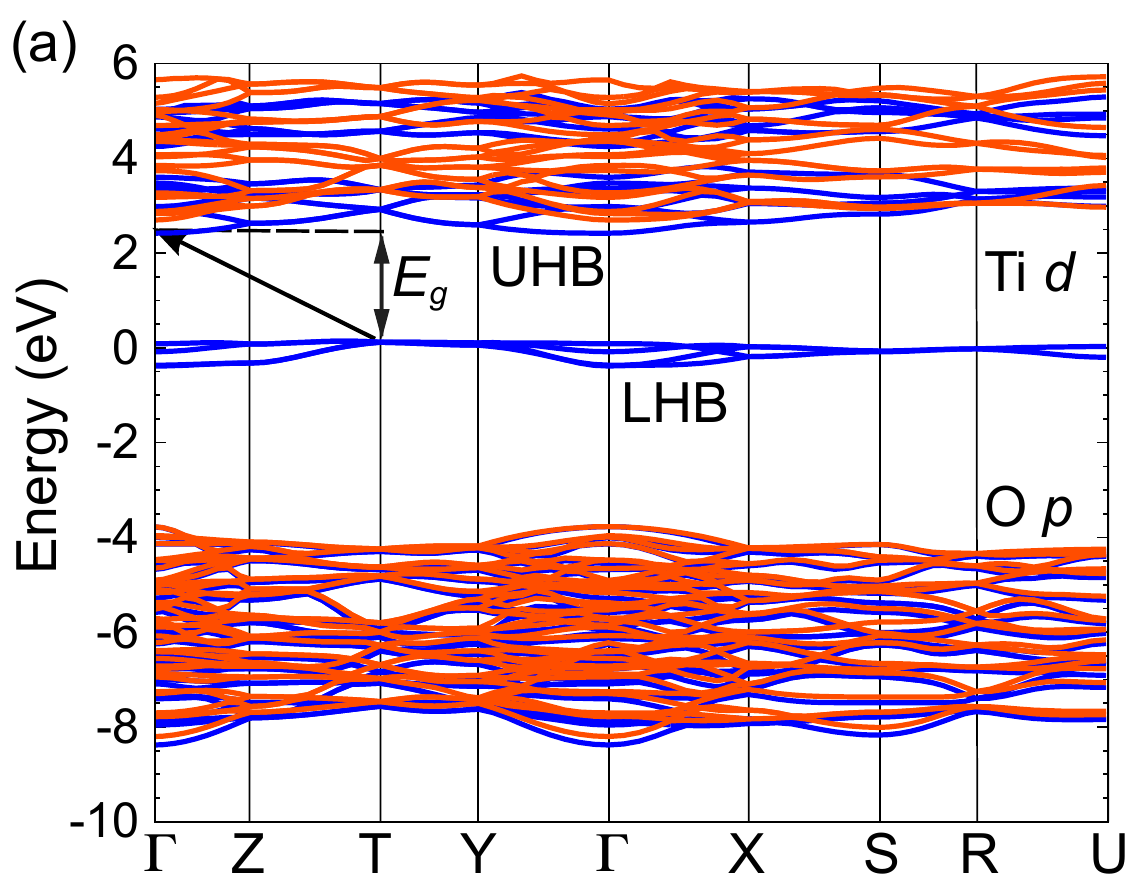}
\includegraphics[width=0.40\textwidth]{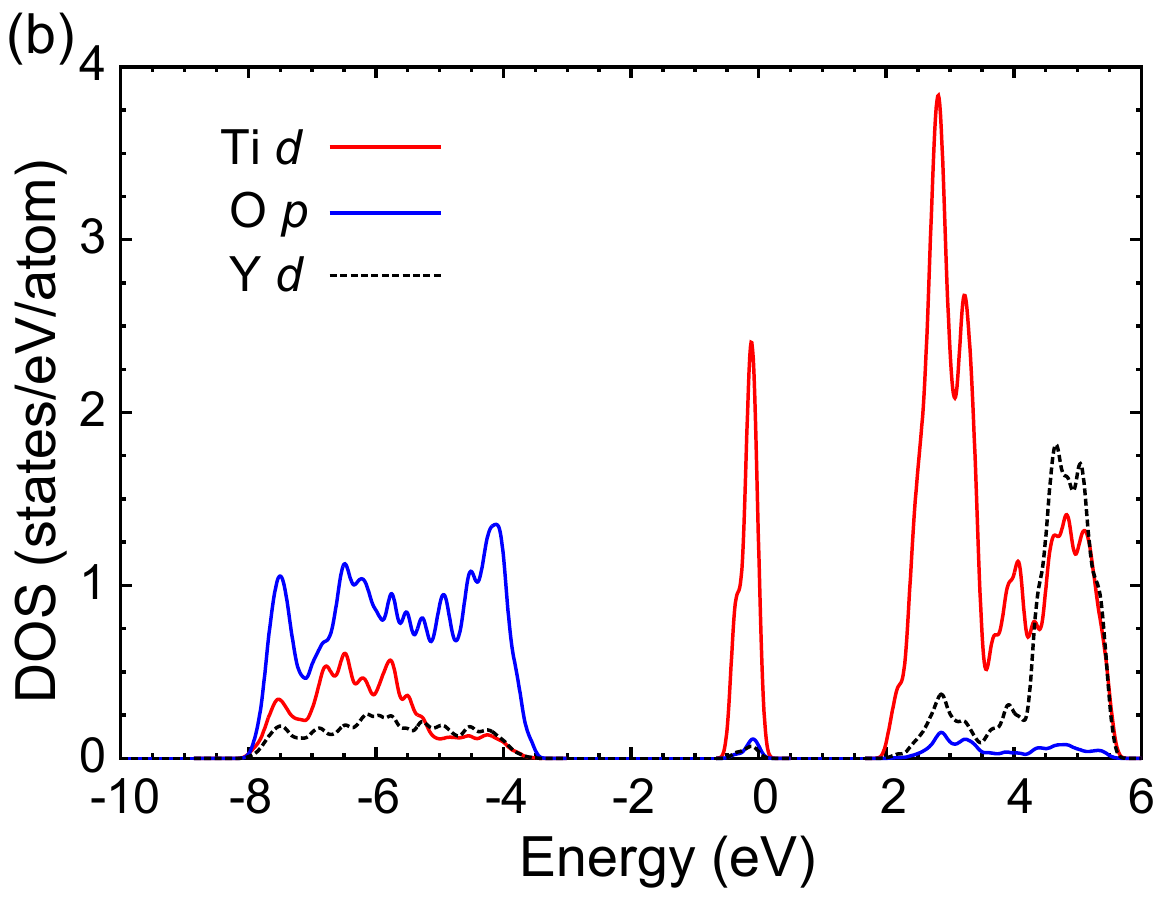}
\caption{\label{fig:hse-bands}(color online) (a) Electronic band structure and (b)
atom-projected density of states (DOS), calculated with the HSE functional.
Spin-up states are represented in blue (darker lines), and spin-down in orange (lighter lines).
The lower and upper Hubbard bands (LHB and UHB) are indicated. The valence-band maximum (i.e., the top of the LHB) is used as the zero of energy.}
\end{figure}

The projected DOS [Fig.~\ref{fig:hse-bands}(b)] shows that the Hubbard bands are derived mainly from Ti $d$ orbitals.
Small contributions from the hybridization with O $p$ and
Y $d$ states are present, as pointed out in Ref.~\cite{pavarini-njp}.
The separation of the peaks in DOS between O-$p$ and Ti-$d$-derived bands in the occupied
manifold (which is around 4 eV) and between Ti-$d$ and Y-$d$-derived bands in the unoccupied
manifold (which is around 2 eV)
is in agreement with the findings of PES and inverse-PES experiments~\cite{arita-ipes,bocquet-pes,morikawa-pes}.

Our calculated band gap $E_g$=2.07 eV is considerably
larger than the optical absorption onset at around 0.6 eV observed in optical
conductivity experiments~\cite{kovaleva-opt,gossling-opt,okimoto-opt}.
From the calculated band structure, interband transitions are expected
to yield an onset at around 2 eV.
This suggests that processes other than interband transitions are relevant.

In our investigations of optical absorption by carriers we focus on the behavior of holes,
since YTiO$_3$ is reported to be unintentionally $p$-type~\cite{goodenough-p}.  We remove an electron from
the top of the valence band in a supercell containing 160 atoms (2$\times$2$\times$2 repetition of the
20-atom primitive cell), and allow the atomic positions to relax (see Sec.\ II of Supplemental Material).
These calculations were performed using a single $k$ point in the integrations over the Brillouin zone. Both the $\Gamma$ and
$(1/4,1/4,1/4)$ points were tested, resulting in total energy differences that are within 0.02 eV.
In the following we report the results using the $(1/4,1/4,1/4)$ point.

\begin{figure}[!ht]
\includegraphics[width=0.45\textwidth]{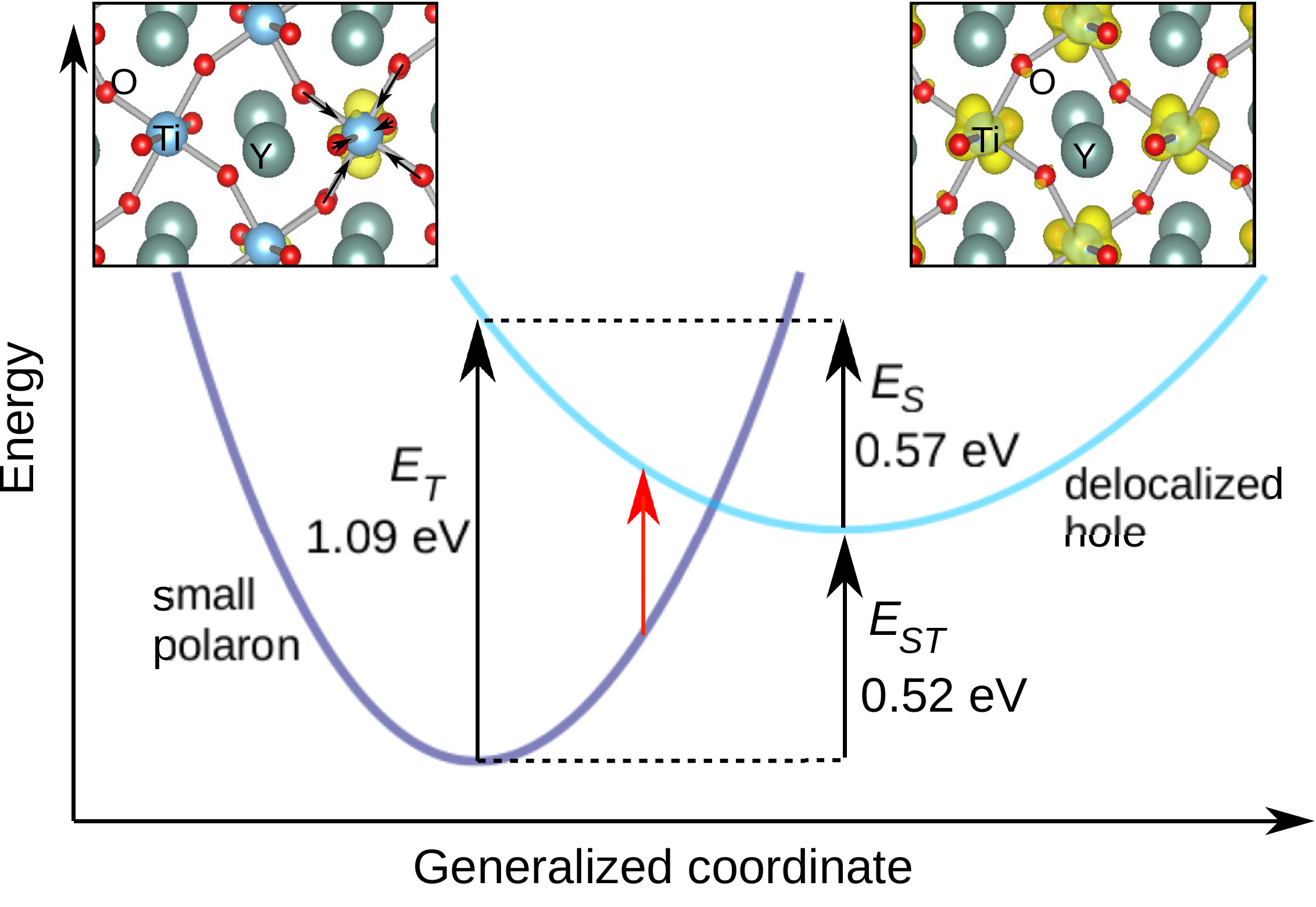}
\caption{\label{fig:conf-diag}(color online) Configuration coordinate diagram for the formation of small
polarons. The insets show the structure and charge density isosurfaces for the small polaron and the delocalized hole.
$E_{ST}$ represents the self-trapping energy of the small polaron, $E_S$ the lattice energy cost,
and $E_T$ the vertical transition energy. The red arrow represents possible lower-energy
excitations. The reported energies result from HSE calculations.
The charge density isosurfaces correspond to the delocalized and self-trapped hole wavefunctions
at 10\% of their maximum value. Arrows in the left inset indicate the lattice distortions around the self-trapped
hole.}
\end{figure}

For a missing electron (one excess hole) in the 160-atom supercell, we find two possible configurations: one corresponding to a delocalized hole,
and another in which the hole is self-trapped.
In the case of a delocalized hole, all the Ti atoms in the supercell contribute equally.
The corresponding charge distribution is extended throughout the crystal (see inset in Fig.~\ref{fig:conf-diag}).
In contrast, the self-trapped hole is localized on a single Ti atom, changing its oxidation state from
Ti$^{+3}$ to Ti$^{+4}$, and is accompanied by a sizable local lattice distortion: the Ti-O bonds contract by 6\%~ of the
equilibrium bond length in HSE, or 4\% in DFT+$U$ (inset in Fig.~\ref{fig:conf-diag}).

We find that the self-trapped hole is significantly more stable than the delocalized hole.  The self-trapping energy of the small polaron ($E_{ST}$),
defined as the energy difference between the delocalized and self-trapped hole~\cite{varley}, is found to be 0.52 eV in HSE calculations.
The localization of the hole leads to an electronic energy gain, but costs lattice energy
(corresponding to $E_{S}$ in the configuration coordinate diagram of Fig.~\ref{fig:conf-diag}).
This cost is computed by evaluating the energy difference between the atomic configuration corresponding to the optimized
atomic coordinates of the small polaron and the configuration corresponding to the delocalized hole in a charge-neutral configuration
(i.e., in the absence of any hole), and is found to be 0.57 eV.
The calculated values of the self-trapping ($E_{ST}$) and strain ($E_{S}$)
energies for the DFT+$U$ and HSE calculations agree to within $0.1$ eV (see Sec.\ II of Supplemental Material).

When small hole polarons are present, electronic transitions can occur from the occupied LHB to the
empty polaronic state.  The vertical excitation energy $E_T$ is the sum of the lattice energy cost $E_{S}$
and the self-trapping energy $E_{ST}$, following the Frank-Condon principle.
The calculated vertical transition energy is 1.09 eV, well below the band-gap energy, and represents a peak in the absorption spectrum.
Furthermore, at finite temperature, vibrational broadening will shift
the absorption onset to lower energies (depicted by the red arrow in Fig.~\ref{fig:conf-diag}).
%
%
We thus propose that small polarons are responsible for the low-energy onset in the optical
conductivity spectra of YTiO$_3$~\cite{kovaleva-opt,gossling-opt,okimoto-opt}.

To obtain a band gap as low as the onset of optical conductivity, one has to
use a value of $U$ substantially smaller than our calculated value. We have found that $U$=1.5 eV
yields a gap of 0.63 eV. A calculation of hole polarons with such a small value of $U$ yields a self-trapping energy
of 20 meV; therefore optical phonons would
easily destabilize the small polarons at finite temperature (see Sec.\ II of Supplemental Material \cite{sup}). Given that the presence of small
polarons is experimentally well established based on hopping conductivity~\cite{goodenough-p}, these results
additionally cast doubt on the interpretation of the 0.6 eV onset as a fundamental gap.

One might think that self-trapped electrons (small electron polarons) could also be important in YTiO$_3$.
However, the formation of a small polaron due to an extra electron on a Ti site in YTiO$_3$ is unfavorable
since the oxidation state Ti$^{+2}$ is unstable against electron delocalization in the conduction band due
to electron-electron repulsion.  We were indeed unable to find localization for an extra electron in the YTiO$_3$ supercell.

As a final point, we address the differences between our results and previous calculations for the electronic structure of YTiO$_3$.
Theoretical studies based on dynamical mean field theory (DMFT)~\cite{pavarini-dmft,craco-dmft},
using a Hubbard $U$ parameter of 5 eV,
yielded a band gap of approximately 1 eV, i.e., 1 eV smaller than our HSE and DFT+$U$ results
\footnote{We stress that the $U$ used in the present work cannot be directly compared to
that of DMFT studies, since it is applied on a different basis set of localized orbitals, and
$U$ depends on the choice of this set~\cite{ucalc-1}}.
These studies were based on an effective Hamiltonian that contains only a subset of the bands
that comprise the LHB and UHB.
While Refs.~\cite{pavarini-dmft,craco-dmft} considered full structural relaxations (at the level of LDA),
they assumed that in YTiO$_3$ the splittings of the $d$ orbitals are
approximately determined by the octahedral crystal field, which would result in
doubly degenerate higher-energy $e_g$ and triply degenerate lower-energy $t_{2g}$ states.
The single electron per Ti site would then justify the use of a Hamiltonian based on $t_{2g}$-derived bands only.

However, in YTiO$_3$, the crystal field does not have a simple octahedral form.
Even in the case of a cubic crystal, with untilted TiO$_6$ octahedra, the Y atoms modify the crystal field
and split the $t_{2g}$ states.
The octahedral rotations break the remaining degeneracies, yielding non-degenerate $d$ orbitals.
Our full band-structure calculations confirm this argument, and reveal that
the LHB contains a large contribution from one of the $e_g$ states.
%
%
\begin{figure}[!ht]
\includegraphics[width=0.45\textwidth]{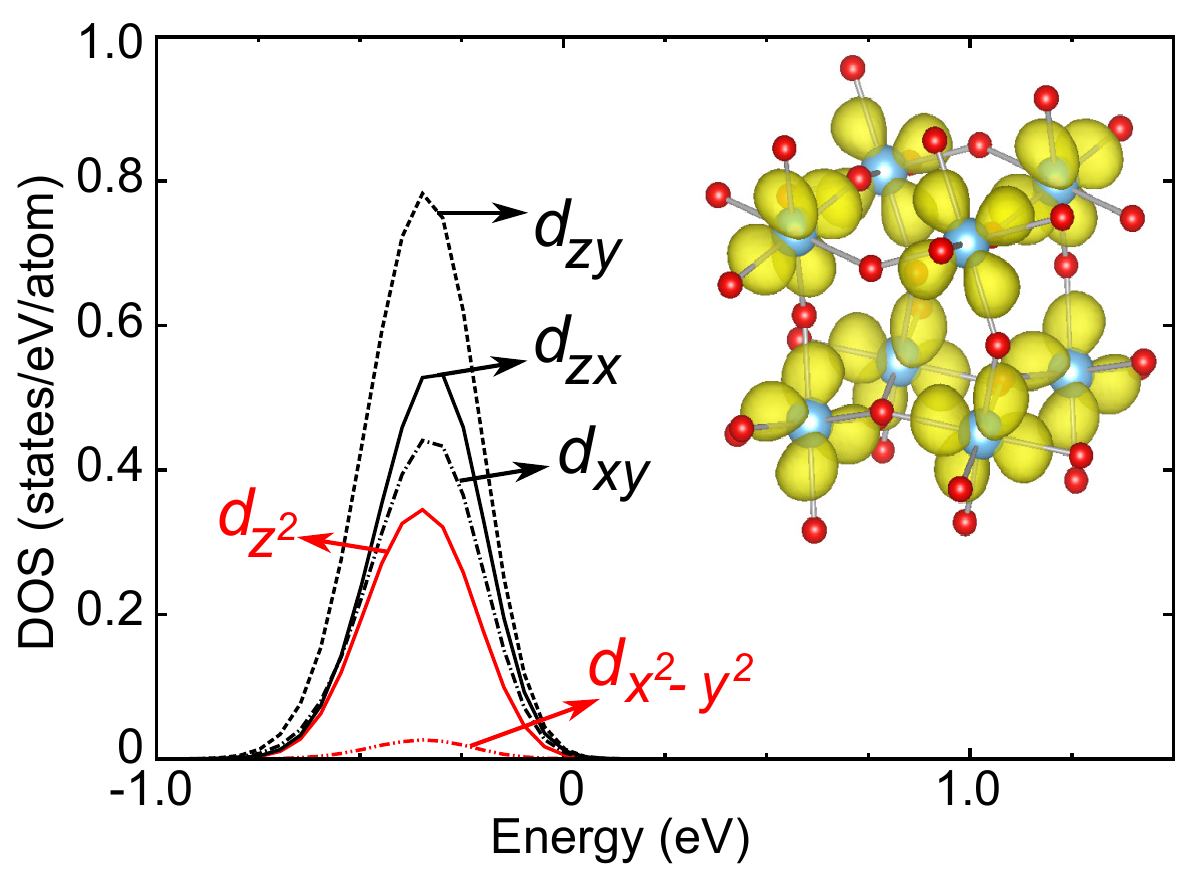}
\caption{\label{fig:lhb}(color online) Density of states (DOS) for
the LHB, obtained in DFT+$U$. The inset shows the total Ti $d$ charge density
associated with the LHB; only O and Ti atoms are shown
for clarity.}
\end{figure}
Figure~\ref{fig:lhb} shows that the LHB is not purely composed of
$d_{zx}$, $d_{zy}$ and $d_{xy}$ (the $t_{2g}$ states), but has important contribution from
other orbitals as well~\footnote{The charge density in the inset of Fig.~\ref{fig:lhb}
corresponds to the highest eigenvalue of the occupation matrices (${\bf n}^{I \sigma}$) that appear
in the DFT+$U$ functional~\cite{ucalc-1}. This highest occupied eigenvalue has a value of $0.96$ and represents,
to a good approximation, the contribution of Ti $d$ states to the LHB.}.
If the Ti $d$ states contributing to the LHB had purely $t_{2g}$ character, the resulting charge density would exclusively exhibit
lobes that point in between the Ti-O bonds.
Instead, the charge density in the inset of Fig.~\ref{fig:lhb} is distinctly nonzero along the Ti-O bonds, providing
clear evidence of contributions from $d_{z^2}$ states.
In fact, the alignment of $d$ orbitals on different Ti sites forming the LHB is in overall agreement with the
orbital polarization observed in Refs.~\cite{pavarini-dmft,craco-dmft}, which is a consequence of octahedral rotations.
The only difference in the orbital polarization picture
is the full inclusion of $e_g$ orbitals.
%


In further quantitative support of this argument, we
have implemented a DFT+$U$ functional that only contains corrections to $t_{2g}$ states in
the PWSCF code,
\begin{equation}
E_U = \frac{U}{2}\, \sum_{I,\sigma}\, {\rm Tr}\, [ {\bf n}_{t_{2g}}^{I \sigma}\, ( 1 - {\bf n}_{t_{2g}}^{I \sigma} ) ],
\end{equation}
with ${\bf n}_{t_{2g}}^{I \sigma}$ being 3$\times$3 matrices, in contrast to
the full corrective functional in Eq.~($9$) of Ref.~\cite{ucalc-1}).
We apply the $U$ correction on the atomic $t_{2g}$ states.
This is not completely equivalent to the approach of Refs.~\cite{pavarini-dmft,craco-dmft},
which was based on Wannier functions starting from $t_{2g}$ projectors and may thus provide some intermixing with $e_g$ and
O $p$ states.
Still, the set of localized orbitals in Refs.~\cite{pavarini-dmft,craco-dmft} is incomplete and thus the corrective $U$
only partially incorporates the missing electronic correlation; in that sense,
our DFT+$U$ calculation with $U$ acting only on $t_{2g}$ states illustrates the effects of using
an incomplete localized orbital basis set in a corrective functional.
Using the same value of the Hubbard $U$ (3.70 eV) as in our full DFT+$U$ calculation, we find that
the gap reduces to 1.25 eV, confirming our argument that inclusion
of {\it all} $d$ orbitals in the corrective treatment (whether DFT+$U$ or DMFT) is essential.
As shown in Ref.~\onlinecite{kotliar-dmft}, DFT+$U$ and DMFT should be in agreement at low temperatures
when $U$/$W$$\gg$1 (i.e., the regime where static correlations dominate), where $U$ is the on-site Coulomb repulsion
and $W$ is the bandwidth associated with the hopping amplitude between $d$ orbitals
of neighboring Ti sites ($t_{dd}$); the latter is negligible in YTiO$_3$.
Note that this bandwidth differs from the actual bandwidth of the LHB, which is determined by
the hopping amplitude from O $p$ to Ti $d$ orbitals ($t_{pd}$).
Furthermore, there are no delocalized electronic states around the Fermi level that
could dynamically screen the effective interaction between electrons residing on Ti $d$ orbitals.
%
%
Thus, the $U$/$W$$\gg$1 condition is satisfied for YTiO$_3$, and static
correlations determine the main features of the electronic structure.
Furthermore, the band gap is not expected to depend very strongly on temperature~\cite{cardona}.
We therefore argue that the gap of $\sim$1 eV observed in DMFT calculations is an artifact of the insufficient
number of bands used in the Hamiltonian.

In summary, we have studied the electronic structure of YTiO$_3$ using DFT+$U$
(with a self-consistently calculated value of $U$) and the HSE hybrid functional.  Both approaches yield a
band structure in agreement with PES/inverse-PES experiments, but with a gap much larger than the experimentally observed
onset of optical conductivity.
We attribute this low-energy absorption to excitations of electrons from the LHB into small-polaron hole states.
The remarkable agreement between the results obtained with the two independent approaches
that include missing static correlations in the exchange-correlation functional strengthens our conclusions.
We have also addressed the discrepancy with earlier calculations that showed a gap around 1 eV,
demonstrating that an incomplete accounting for electronic localization leads to in an underestimation of the band gap.
Our qualitative conclusions likely apply to closely related systems such as LaTiO$_3$, GdTiO$_3$, and
possibly vanadates with $3d^1$ orbital configuration, in which similar confusion regarding the band gap and
possible misinterpretations of experiment are common.
A correct assessment of the electronic band gap has profound implications for the design of
interfaces and devices based on rare-earth titanates and related oxides.

This work was supported by the MURI program of the Office of Naval Research, Grant Number N00014-12-1-0976.
Additional support was provided by ARO (W911-NF-11-1-0232) and by the NSF MRSEC Program (DMR-1121053).
Computational resources were provided by the Center for Scientific Computing at the CNSI and MRL (an NSF MRSEC, DMR-1121053) (NSF CNS-0960316), and by the Extreme Science and Engineering Discovery Environment (XSEDE), supported by NSF (OCI-1053575 and DMR07-0072N).

\bibliography{hjvdw}

\end{document}



\title{Supplemental Material for\\
Interband and polaronic excitations in YTiO$_3$ from first principles}

\author{Burak Himmetoglu, Anderson Janotti, Lars Bjaalie, and Chris G. Van de Walle}

\maketitle

\section{DFT+$U$ band structure and the effect of Hund's coupling $J$}

\begin{figure}[!ht]
\includegraphics[width=0.40\textwidth]{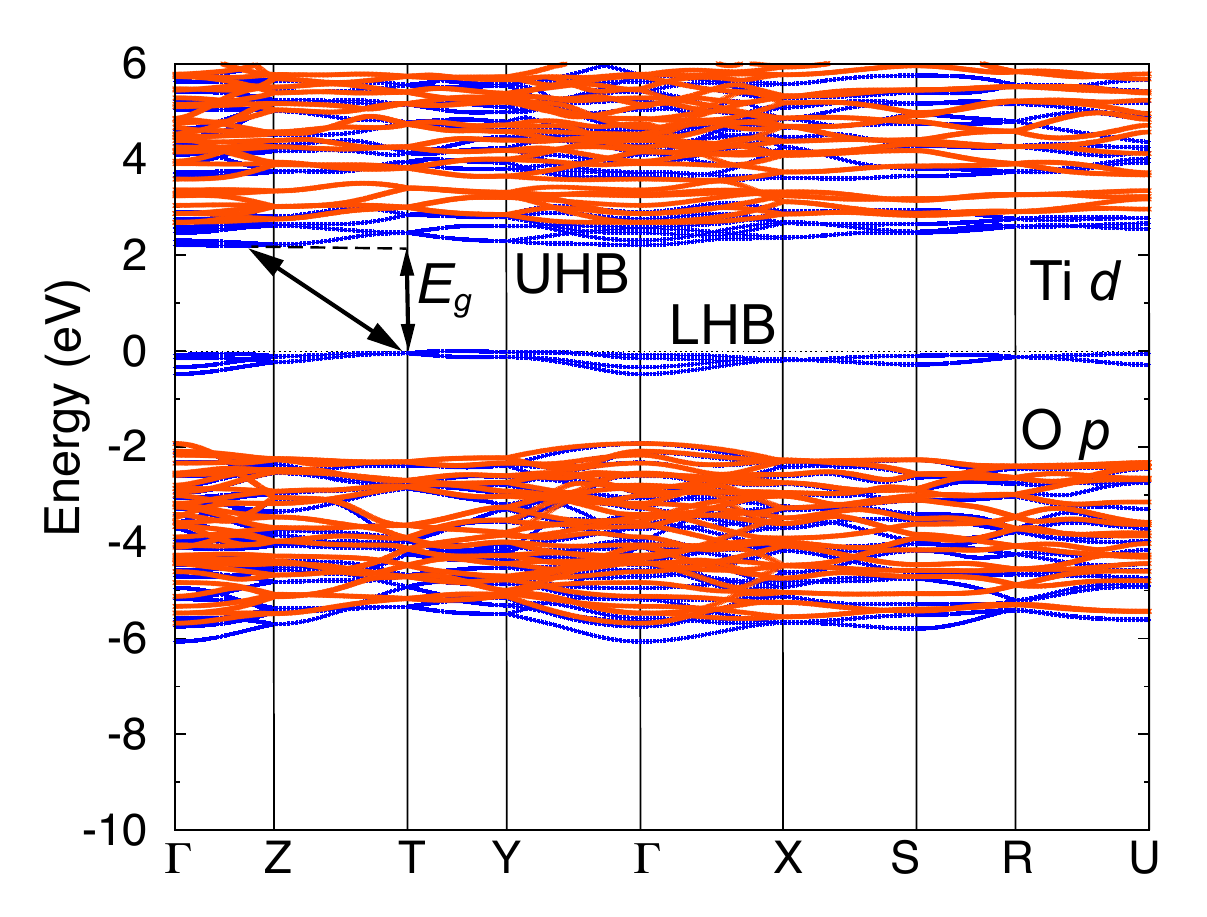}
\includegraphics[width=0.40\textwidth]{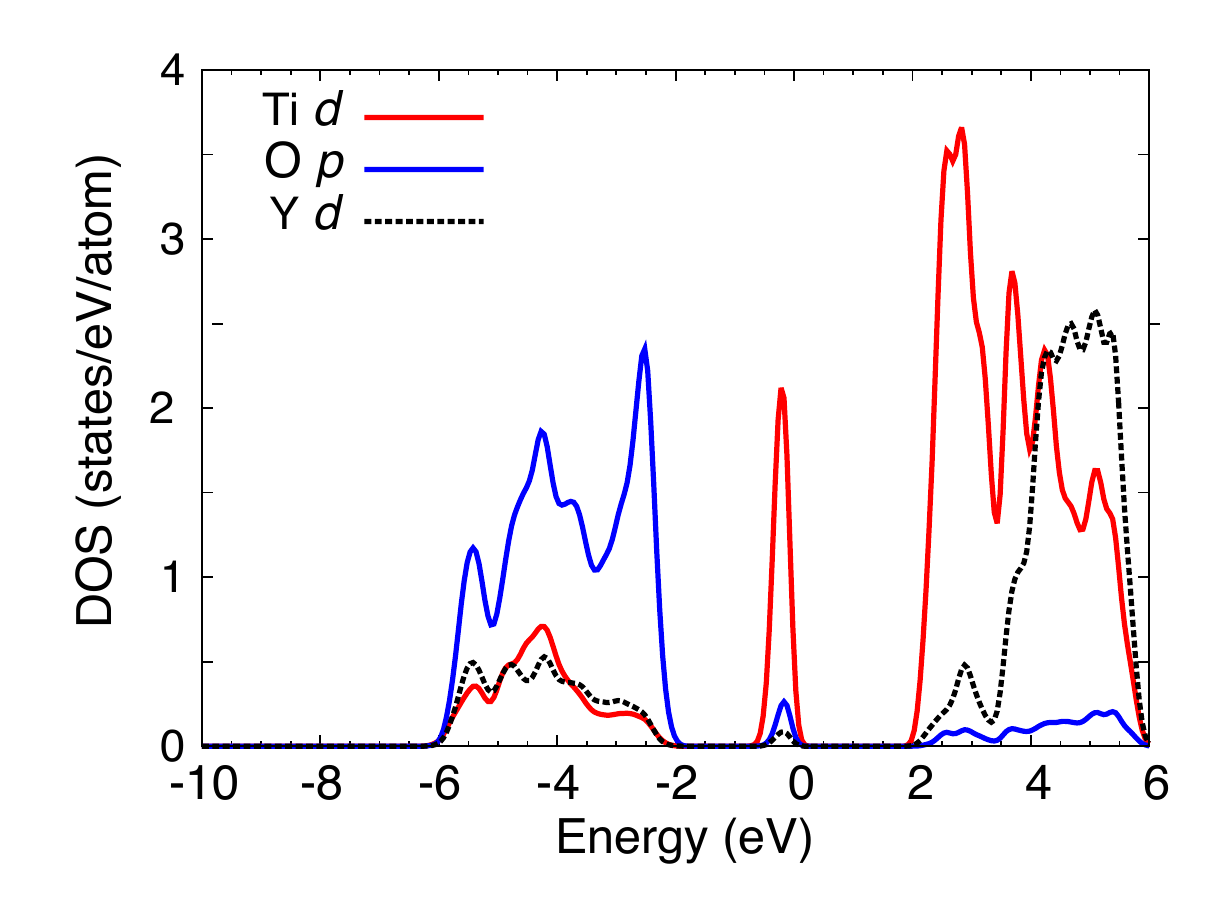}
\caption{\label{fig:dftu-bands}((a) Electronic band structure and (b)
atom-projected density of states (DOS), calculated with DFT+$U$.
Spin-up states are represented in blue (darker lines), and spin-down in orange (lighter lines).
The lower and upper Hubbard bands (LHB and UHB) are indicated. The valence-band maximum (i.e., the top of the LHB) is used as the zero of energy.}
\end{figure}

The DFT+$U$ ($U$=3.70 eV) band structure and projected DOS are shown in Fig.~\ref{fig:dftu-bands},
in a format to be directly compared with Fig.~2 of the main text.
The valence-band maximum is located between Z and T and the
conduction-band minimum is between $\Gamma$ and Z points. The band gap is 2.20 eV,
only slightly larger than the HSE gap of 2.07 eV.
The overall band structures obtained with DFT+$U$ and HSE are in good agreement with each other, particularly
for the $d$-orbital derived bands that are at the core of our study.
Some differences occur for the bands derived from O $p$ states,
which are around 2 eV closer to the LHB in DFT+$U$ than in HSE.
This is an expected result, since DFT+$U$ only acts on
the bands derived from Ti $d$ states, and does not produce significant corrections to others.

We have also considered the effect of Hund's coupling $J$ in the DFT+$U$ scheme.
The on-site parameters $U$ and $J$ were self-consistently computed using the linear response method~\cite{ucalc-1,ucalc-2,ucalc-3},
resulting in $U$=4.24 eV and $J$=1.04 eV. We note that this $U$ value
is slightly different from the value reported in the main text, since the self-consistency procedure of updating
$U$ after each linear-response calculation is now based on the DFT+$U$+$J$ wavefunction rather than just DFT+$U$.
The resulting ground state yields a band gap of 1.91 eV, which is still
close to the HSE (2.07 eV) and DFT+$U$ (2.20 eV) results, and significantly larger than the onset of optical conductivity.

The fact that inclusion of Hund's coupling $J$ has only a minor effect can be attributed to the specific orbital occupation.
Indeed, $J$ plays an important role in controlling magnetic ordering if more than one electron resides on the same orbital
manifold.  For the $d^1$ orbital occupation in YTiO$_3$, however, no significant effects are expected, and are indeed not observed.

\begin{figure}[!ht]
\includegraphics[width=0.40\textwidth]{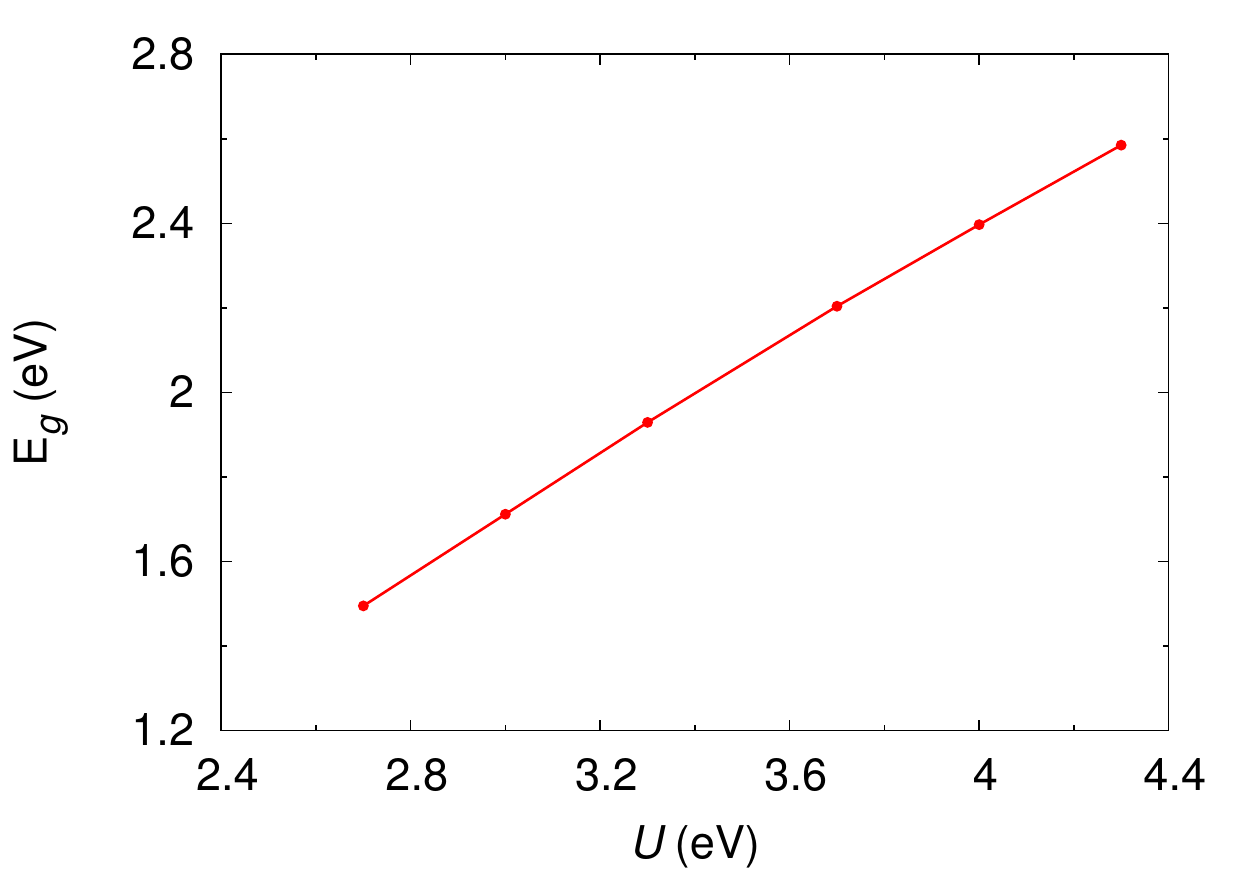}
\caption{\label{fig:egvsu}Variation of the band gap as a function of $U$.}
\end{figure}

We have also investigated the variation of the band gap as a function of $U$ in Fig.~\ref{fig:egvsu}.
In the procedure for calculating $U$,
self-consistency is assumed when the difference between calculated values of $U$ between iterations
is smaller than 0.1 eV. The actual accuracy of the $U$ calculation may be lower due to numerical
errors inherent to the linear response approach.
A conservative upper bound for the error on $U$ can be obtained as $U^2\, \alpha$ [Eq.(19) of Ref.~\onlinecite{ucalc-1}), where
$\alpha$ is the potential shift applied on the $d$ orbitals for constructing the linear-response
matrices. This yields an error $\vert \Delta U \vert <$ 0.6 eV.
As can be seen in Fig.~\ref{fig:egvsu}, even with a variation of $\pm$ 0.6 eV the resulting band
gap is still much larger than the onset of optical conductivity at 0.6 eV reported in experiments.
In order to obtain a band gap of 0.6 eV, a value of $U$ as low as 1.5 eV would need to be employed.

\section{Small polarons in DFT+$U$ and HSE}

\begin{figure}[!ht]
\includegraphics[width=0.15\textwidth]{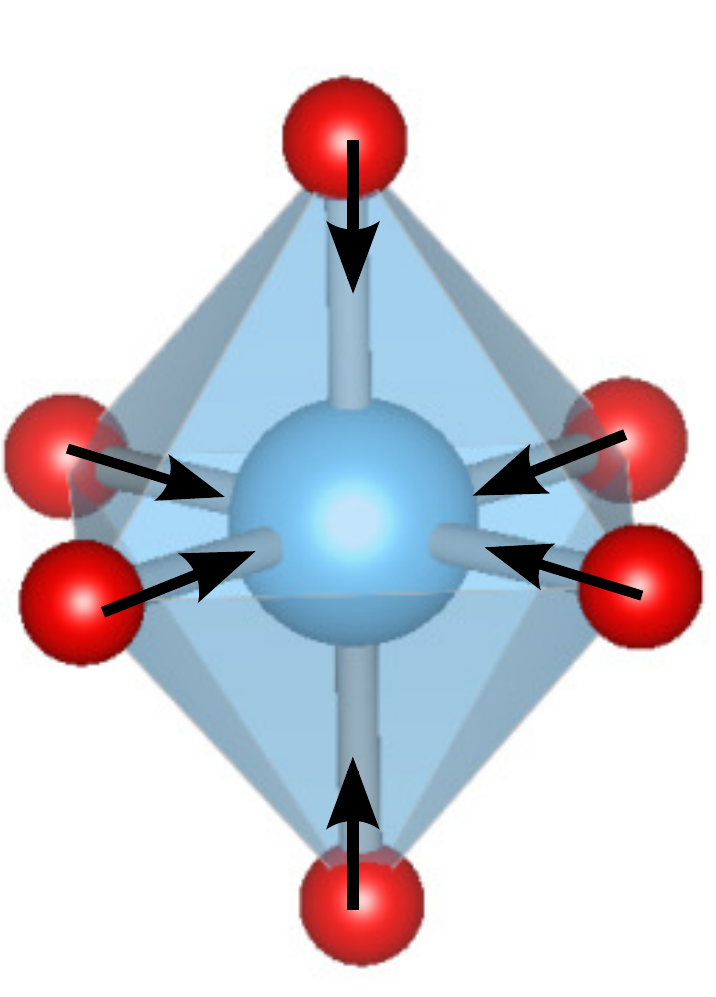}
\caption{\label{fig:octa}Oxygen atom distortions around the Ti atom where self-trapping of a hole
is obtained.}
\end{figure}

In our calculations, the small-polaron configuration is studied by providing an initial inward perturbation to the O atoms around
a randomly chosen Ti atom within the $2 \times 2 \times 2$ supercell, as shown in Fig.~\ref{fig:octa}, with rotated octahedra.
One electron is removed from the system, and the magnetization of the central Ti atom is initially set to zero.
Charge neutrality is provided by a homogeneous background.
Structural optimization with these initial
conditions leads to the stabilization of the small-polaron state. The delocalized hole solution, on the other hand,
results from initial conditions when no inward distortions are present.

In Table~\ref{tab:pol} we report the vertical transition energy ($E_{T}$), self-trapping energy ($E_{ST}$) and
the crystal strain energy ($E_{S}$) calculated using DFT+$U$ and HSE. As discussed in the main text,
both methods yield results in close agreement with each other.
\begin{table}[!ht]
\caption{\label{tab:pol} Vertical transition energy ($E_T$), self-trapping energy ($E_{ST}$)
and crystal strain energy ($E_S$) calculated with DFT+$U$ and HSE.}
\begin{ruledtabular}
\begin{tabular}{cccc}
  & $E_T$ (eV) & $E_{ST}$ (eV) & $E_{S}$  (eV)\\
\hline
DFT+$U$  & 1.21 & 0.63 & 0.58 \\
HSE    & 1.09 & 0.57 & 0.52 \\
\end{tabular}
\end{ruledtabular}
\end{table}

As noted in Sec.\ I, if the value of $U$ were chosen to be 1.50 eV, DFT+$U$ calculations
would yield a band gap around 0.6 eV (0.63 eV, to be precise), close to the onset of optical
conductivity observed in experiments.
It is informative to calculate self-trapping and crystal strain energies for this smaller $U$ value.
With $U$=1.50 eV,  we find that $E_{ST}$=20 meV and $E_{S}$= 0.29 eV.
This shows that for $U$ values consistent with a 0.6 eV gap, the small polaron would not be stable at finite
temperatures, since it can be destabilized by optical phonons whose energies range between 18 and 72 meV~\cite{yto-op}.
This is inconsistent with resistivity data which clearly show hopping conductivity due to self-trapped holes~\cite{goodenough-p}.
We conclude that interpreting the onset of conductivity at 0.6 eV as the signature of the band gap (which would require $U$=1.50 eV in the calculations) cannot be reconciled with the observation of small polarons in the material. This provides additional support for our results in which physical values of $U$ (close to 4 eV) yield a large band gap as well as stable hole polarons.
